\documentclass[useAMS,usenatbib]{mn2e}
\usepackage{graphicx} 

\title[H$\alpha$ Surface Brightness in Outer Disk]{A Spectroscopic Study of the H$\alpha$ Surface Brightness Profiles in the Outer Disks of Galaxies}

\author[Christlein, Zaritsky \& Bland-Hawthorn]{Daniel Christlein\\Max-Planck-Institut f\"ur Astrophysik, Garching, Germany
\newauthor Dennis Zaritsky \\ Steward Observatory, University of Arizona, Tucson, AZ, U.S.A.
\newauthor Joss Bland-Hawthorn \\ Sydney Institute for Astronomy, School of Physics A29, University of Sydney, Australia}

\begin{document}
\maketitle

\begin{abstract}         
The surface brightness profile of H$\alpha$ emission in galaxies is generally thought to be confined by a sharp truncation, sometimes speculated to coincide with a star formation threshold. 
Over the past years, 
observational evidence for both old and young stellar populations, as well as individual H II regions, 
has demonstrated that the outer disk is an actively evolving part of a galaxy. To provide constraints on the origin of the aforementioned H$\alpha$ truncation and the relation of H$\alpha$ emission in the outer disk to the 
underlying stellar population, we 
measure the shape of the outer H$\alpha$ surface brightness profile of 15 isolated, edge-on late-type disk galaxies 
using deep, long-slit spectroscopy. Tracing H$\alpha$ emission 
up to 50\% beyond the optical radius, $R_{25}$, we find
a composite H$\alpha$ surface brightness profile,
well described by a broken-exponential law, 
that drops more steeply in the outer disk, but which is not truncated.
The stellar continuum and H$\alpha$ surface brightness both exhibit a break at $\sim0.7 R_{25}$, but the H$\alpha$ emission drops more steeply than the stellar continuum beyond that break. Although profiles with truncations or single exponential laws correctly describe the H$\alpha$ surface brightness profiles of some individual galaxies, flexible broken-exponentials are 
required in most cases and are therefore the more appropriate generic description.
The common existence of a significant second surface brightness component beyond the H$\alpha$ break radius disfavors the hypothesis that this break is a purely stochastic effect.
\end{abstract}

\begin{keywords}
galaxies: kinematics and dynamics -- galaxies: formation -- galaxies: evolution -- galaxies: spiral -- galaxies: structure
\end{keywords}

\section{Introduction}
\label{sec:intro}

The outer disks of galaxies are increasingly being recognized as a
diverse environment whose properties may aid us in understanding the
formation of galaxies. Since the radio observations of the 1950s  \citep{vandehulst,dieter}, we have known that the baryonic component of
spiral galaxies, as traced by neutral hydrogen, is significantly larger than suggested by classical
size indicators such as $R_{25}$ (the radius where the surface brightness drops to 25 mag arcsec$^{-2}$; we use measurements of the $B$-band surface brightness obtained from the literature throughout this analysis).  Optical studies, however, have focused on the inner parts of galaxies, where, among other aspects, the radial distribution of H$\alpha$ emission has been subject of intense scrutiny for many years. For example, \citet{martinkennicutt} used narrow-band imaging of nearby galaxies to construct H$\alpha$ surface brightness profiles.
Such profiles usually exhibit a ``truncation" in the H$\alpha$ surface brightness at relatively large radii, although usually still within the optical radius $R_{25}$. This truncation is also reflected in the scarcity of H$\alpha$ rotation curves that extend to radii much beyond $\sim 0.7 R_{25}$ in comprehensive long-slit surveys (e.g., \citet{vogt}, being one of the deepest such surveys, observe no rotation curves beyond $R_{25}$ in a sample of 329). A widely accepted hypothesis for such a
truncation is the existence of a threshold in
the surface gas density, below which star formation becomes
inefficient \citep{kennicutt89,martinkennicutt}. Alternate explanations include the hypothesis
that the truncation is indicative of an actual break in the mass
distribution that is related to the initial formation conditions of the disk \citep{vanderkruit87}.

However, ``truncation" may be too strong a descriptive statement because
it is now known that low-level star formation occurs at 
large galactocentric radii, well beyond the radius corresponding to the critical threshold. \citet{blandhawthorn,ferguson1998a,ferguson1998b} have detected
individual H$\alpha$ emission regions in the outer disks of spiral
galaxies as far as 2 $R_{25}$, and UV emission appears relatively common in outer disks \citep{thilker,zaritskychristlein,thilker2}. Furthermore, in a study of NGC 3814 using deep two-band optical imaging, \citet{herbertfort} have found statistically significant overdensities of marginally resolved sources in the outer disk that are likely to be star clusters.

This outer-disk star formation has raised new interest in the hypothesis of a star formation threshold. The absence of evidence for a break in the surface brightness profiles of UV emission, which is also a star formation indicator, led
\citet{boissier,boissier2} to suggest that the star formation rate does not exhibit a break, and that the H$\alpha$ truncation is a stochastic effect. As the expectation number of star formation regions with stars massive enough to generate a Stroemgren sphere drops below unity, the H$\alpha$ emission profile goes to zero, while low level(mass) SF continues as measured with the UV.  Other, perhaps more exciting, possible explanations include a change in the initial mass function \citep{meurer}. These arguments depend critically on reliable measures of H$\alpha$ and continuum profiles out to large radii.

At the same time, studies of faint optical continuum emission have shown that stellar disks can also be traced
far into the outer disk. While some galaxies exhibit a single exponential surface brightness profile to the largest measurable radii \citep{blandhawthorn2005}, others are more accurately described by broken exponential profiles with a characteristic break radius, beyond which the stellar surface brightness profile may be shallower or steeper than in the inner disk \citep{pohlen,erwin}. These profiles 
are sometimes described as exponential, sub-exponential, and super-exponential \citep{vlajic2009}; the exponential and sub-exponential types are also variously referred to as Freeman Types I and II \citep{freeman}, and the designation ``Type III" has come into usage to describe the super-exponential (up-bending) shape.

How these outer stellar structures are related to possible structures in the gaseous disk, and in particular, H$\alpha$ emission, is a crucial question in understanding the origin of the outer disk stellar populations.  Is the truncation in H$\alpha$ emission a real indicator of a truncation in star formation, or, as \citet{boissier,boissier2} suggest, merely a stochastic effect? If, on the other hand, the break in the star formation surface density is real, is it also responsible for the characteristic break in the stellar continuum profile? In that case, is there significant star formation beyond the break, and is it enough to have created the outer disk stellar content in situ? Or must other processes be invoked to populate the outer disks with stars?

To answer these questions, a systematic and quantitative study of the distribution of H$\alpha$ emission at large radii is required. The traditional technique of narrow-band imaging with subsequent subtraction of broad-band continuum emission is generally insufficiently sensitive to probe to large radii. The limitation lies not only in the achievable signal-to-noise ratio (which is limited because typical narrow-band filters are much wider than the H$\alpha$ emission line), but also in the stellar continuum subtraction. At large radii, because the spectral flux density at the peak of the H$\alpha$ line is of the same order of magnitude as the stellar continuum, the wavelength dependence of the stellar continuum spectral energy is sufficient to 
significantly degrade the subtraction. For example, absorption troughs, which the H$\alpha$ emission line is often embedded in, can render it undetectable. 

The way around these problems is to use higher spectral resolution. This can involve
very narrow filter bandpasses, which is now possible over a wide range of redshifts due to the
increasing availability of tunable filters
\citep{bh98,cepa}, or traditional spectroscopy.
In a pioneering effort  in the 
study of gaseous outer disks, \citet{blandhawthorn} detected H$\alpha$
emission at $\sim 1.25$ R$_{25}$, beyond the truncation radius of
the neutral hydrogen disk, using the Fabry-Perot staring technique.
However, true spectroscopic observations provide improvements in the
subtraction of the stellar
continuum, of the [NII] emission lines, and of H$\alpha$ absorption
troughs, 
allowing one to reach sensitivity limits fainter than 10$^{-18}$ erg s$^{-1}$
cm$^{-2}$ arcsec$^{-2}$ (see also \citet{madsen}). 

We have observed
late-type, edge-on disk galaxies with multi-hour, long-slit
spectroscopy. The long-slit technique, although suffering from the
much lower throughput in comparison to the Fabry-Perot 
technique, allows us, for edge-on geometry, 
to take advantage of fields of view that cover
the entire galaxy and greatly facilitates sky and continuum subtraction. 
We typically detect H$\alpha$ emission to galactocentric
radii of 1.5 R$_{25}$, and in certain 
cases up to 2 R$_{25}$. 
In a previous paper \citep{christleinzaritsky}, we discuss the kinematic properties of these outer H$\alpha$ disks and find the kinematics to be disk-like, with generally no indication of kinematic anomalies or higher velocity dispersions as one approaches the outer edge of the disk.
A subsample of galaxies with known optical warps has been studied separately \citep{christleinbh} to determine if kinematic anomalies, such as breaks in the rotation curve, were associated with the onset of warps that could indicate ongoing accretion processes of compact HI clouds or satellite galaxies as causes of the warp. We use data from both of these samples in this paper.

We now turn our attention to the surface brightness profile of H$\alpha$ emission in the outer disk. Our aim is to measure the distribution of H$\alpha$ emission in the outer disk and determine if the H$\alpha$ emission profiles can be categorized as done for the stellar continua. We will determine whether there is indeed a break in the H$\alpha$ surface brightness profile, and if so, whether this break can be associated with that in the stellar surface brightness. If there is a break, is it steep enough to constitute a truncation? And how is the stellar continuum, a measure of integrated star formation, distributed in comparison to the current star formation, as indicated by H$\alpha$?

We describe our sample in \S \ref{sec_data}. 
In \S \ref{sec_composite}, we discuss and compare the properties of H$\alpha$ and stellar continuum emission based on a composite of our 15 individual galaxy spectra, while in \S \ref{sec_individual}, we discuss what constraints we can place on the profile shapes of individual galaxies.

\section{Data}
\label{sec_data}

Our sample comes from three observing runs: a three-night run using FORS1 at the VLT in November 2004, a three-night run with GMOS at Gemini-South in April 2005, and a three-night run using FORS2 at the VLT in September 2007. Target selection was, in all cases, for isolated, edge-on, late-type galaxies of several arcminutes in diameter, preferentially at a redshift of several thousand km/s, which places the H$\alpha$ line in a relatively quiet window of the sky background. An important difference among the runs is that the targets for the first two runs were chosen to be as morphologically undisturbed and regular as possible, whereas targets for the third run deliberately included several objects with optical warps, mostly taken from the catalog by \citet{sanchezsaavedra}. 
At first, it may seem incongruous to include a set of warped galaxies in a long-slit experiment, because the disk material is eventually going to curve away from the slit at the onset of the warp. However, given that nearly all galaxies might be warped to some extend \citep{sanchezsaavedra}, even if they appear ``normal" in the available imaging, it is useful to include a subsample of known warped galaxies to provide the internal control against which to view our results. Furthermore, warps typically set in around the $R_{25}$ radius \citep{briggs}, so that several important results from the undisturbed sample within this radial distance may still be verified with the warped sample.

All 15 targets are listed in Table \ref{tab_sample}. The Table lists
the R$_{25}$ radii adopted for this analysis, summarizes which telescope and instrument were used, and identifies whether the galaxy is known to be warped or not.  Values of R$_{25}$ are drawn from \citep{esolv} for all objects from the FORS1 and GMOS-S runs and from the Third Reference Catalogue of Bright Galaxies \citep{devauc} for all objects from the FORS2 run. The projected optical diameter, $2R_{25}$, of all galaxies is at most
half the slit length, so that each object is covered completely by the
slit in a single observation, along with a substantial sky sample. In our analysis, we group the galaxies into an undisturbed and a warped sample. Two of the "unwarped" galaxies (ESO 340-G026, MCG -01-10-035) were observed during the September 2007 run, which was primarily targeting warped objects.

We describe the surface brightness profile as a
function of projected radial distance from the galaxy center in units
of the R$_{25}$ radius so as to permit a comparison between
galaxies of different physical sizes. Other characteristic size
measures, such as the half-light radius, $R_{50}$,
and the radius containing 90\% of the total
light, $R_{90}$, were also considered. However, all of these are generally strongly
correlated (for all galaxies in our sample and in \citet{esolv}, R$_{25}\approx$R$_{90}$ with a scatter of 15$\%$, and with the exception of a single galaxy, R$_{25}\approx
2\times$R$_{50}$ to within 15$\%$). We choose to use R$_{25}$ because it is 
the only such quantity available in the literature for all of our galaxies and because it is a robust measure for the "edge" of the disk in this sample where inner-disk size measures can be heavily affected by dust.

\begin{table*}
 %\centering
 %\begin{minipage}{200mm}
  \begin{tabular}{llllll}
  \hline
Galaxy & R$_{25}$ [$^{\prime\prime}$] & Run & R$_{lim}^{a} [R_{25}]$ & R$_{lim}^{b} [R_{25}]$ & Warp? \\
 \hline
ESO 201- G 022 & $76^{a}$ & FORS1/VLT & 1.28 & 1.29&N\\ % internal No. 8
ESO 299- G 018 & $62^{a}$  & FORS1/VLT & 1.41 & 1.43&N\\ % internal No. 9
ESO 323- G 033 & $50^{b}$ & GMOS-S/Gemini-S & 1.47 & 1.48&N \\ % internal No. 10
ESO 380- G 023 & $42^{a}$  & GMOS-S/Gemini-S & 0.84 & 0.87&N\\ % internal No. 11
ESO 385- G 008 & $62^{a}$  & GMOS-S/Gemini-S & 1.08 & 1.11&N\\ % internal No. 13
ESO 478- G 011 & $42^{a}$  & FORS1/VLT& 1.45 & 1.34&N\\ % internal No. 14
ESO 340- G 026 & $61^{c} $ & FORS2/VLT & 1.44   & 1.45 & N\\  % internal No. 2
IC 2058 & $93^{a}$  & FORS1/VLT& 1.03 & 1.05&N\\ % internal No. 15
IC 4393 & $80^{a}$  & GMOS-S/Gemini-S& 0.91 & 0.92&N\\ % internal No. 16
MCG -01-10-035 & $70^{c}$ & FORS2/VLT &  1.47 & 1.49&N\\  % internal No. 3
ESO 184- G 063 & $61^{c}$ & FORS2/VLT &  1.44  & 1.44 & Y\\  % internal No. 1
ESO 473- G 025 & $77^{c}$ & FORS2/VLT &  1.02  & 1.03 &Y \\   % internal No. 3
NGC 259 & $85^{c}$ & FORS2/VLT &  1.35 & 1.37 &Y\\ % internal No. 5
UGC 12423 & $104^{c}$ & FORS2/VLT &  1.43 & 1.45&Y\\ % internal No. 6
UGCA 23 & $85^{c}$ & FORS2/VLT &          0.97 & 0.99&Y \\ % internal No. 7
\hline
\end{tabular}\\

\caption{Target list: R$_{25}$ is the major axis radius of the 25 mag arcsec$^{-2}$ isophote ((a): drawn from ESO-LV catalogue; (b): estimate based on acquisition image; close to RC3 value; (c): drawn from RC3). R$^a_{lim}$ and R$^{b}_{lim}$ are the outermost radii at which we detect H$\alpha$, according to the two definitions described in the text.
 }
\label{tab_sample}
\label{tab_outermost}
%
%\end{minipage}
\end{table*}

\subsection{Data Reduction}

All raw spectra are processed using a standard IRAF pipeline. Flux calibration is carried out with spectrophotometric standards taken each night; calibration factors for nights in a given run are consistent within 10\%. Cosmic rays are identified by comparing a pixel value to the average of its neighbours within an annulus at distances between 3 and 10 pixels and to the standard deviation in this annulus. The locus of pixels to be flagged as cosmic rays, depending on these quantities, is determined by visual inspection in a plot of the average neighbour pixel value versus the number of standard deviations. We subtract the sky background and continuum emission along the dispersion axis using the appropriate IRAF tasks.

For each galaxy, we then extract kinematic data, using our own software. Line centroiding is carried out and verified interactively for each individual row on
the CCD. We fit and correct for the local H$\alpha$ stellar absorption troughs, which are not removed by the sky subtraction or polynomial stellar continuum subtraction, with Gaussian
functions (as now typically done
for emission line measurements, see \cite{moustakas}), verify, and, if necessary, adjust the fit manually. We then transform all spectra onto a
common coordinate frame, consisting of the projected separation from
the galaxy centroid in units of R$_{25}$ and the wavelength offset
from the interpolated wavelength corresponding to the mean H$\alpha$ rotation velocity
(i.e., we remove the signature of rotation,  for an example see Figure 8 in \cite{christleinzaritsky}).
The transforming process does adversely affect the final signal-to-noise ratio, but has the advantage that all individual spectra can be compared on a consistent basis despite being taken with three different instruments and exhibiting different rotation curves.

From the transformed spectrum, we measure the H$\alpha$ flux within a tophat kernel that generously encompasses the entire line, including its deviations from the interpolated rotation curve. The flux is determined by integrating the counts across the entire kernel and renormalizing  to units of erg s$^{-1}$ cm$^{-2}$ arcsec$^{-2}$. 
We determine uncertainties in the extracted flux by sliding the
same tophat kernel 
along the wavelength axis in a ``background" region without strong
emission or absorption lines.
We finally determine the standard deviation of this residual flux over all background positions and adopt that value as the uncertainty in the background. We do not specifically include a term to account for Poisson uncertainties in the emission line flux, because our
analysis focuses on a low-surface brightness regime where background
errors are dominant.

We calculate the stellar continuum brightness, using the same spectra without the stellar continuum subtraction, from the mean of
background regions that lie adjacent to the H$\alpha$ line 
and which are free of any strong emission or absorption features. The uncertainty in this
quantity is calculated as the error of the mean from the scatter of
the individual pixel values. 

\section{Results}

\subsection{Overview}
\label{sec_overview}

Before we begin our discussion, we set the stage by discussing two key considerations.
First,  we need to define a set of models that we will test against the data. 
Our aim is to characterize the surface brightness profile of the H$\alpha$ and stellar continuum emission quantitatively. For this purpose, we consider three basic hypotheses:
\begin{itemize}
\item $H_1$: A broken exponential profile characterized by the equation
\begin{equation}
f(r) = c_0  e^{-\lambda_1  r} (1-\Theta(r-r_0)) + c_0  e^{-\lambda_1 r_0} e^{-\lambda_2 (r-r_0)} \Theta (r-r_0).
\label{eq_brokenexp}
\end{equation} 
This is the most flexible model. It consists of two independent, but
contiguous exponential profiles with a break radius of $r_0$,
characterized by a slope $\lambda_1$ in the inner, and $\lambda_2$ in the outer
disk. The normalization is adjusted by the constant $c_0$. The
function $\Theta$ is the step function, i.e., $\Theta(x)=1$ for $x\geq0$,
and $0$ otherwise. 
\item $H_2$: A truncated exponential profile characterized by
\begin{equation}
f(r) = c_0  e^{-\lambda_1  r} (1-\Theta(r-r_0))  .
\end{equation}
This surface brightness profile has a sharp cutoff at radius $r_0$ and is zero at larger radii. Free parameters are $c_0$, $\lambda_1$, and $r_0$.  A sharp truncation is an idealized scenario, and even if it applied to the radial surface brightness profile, projection effects in a nearly edge-on sample would soften that edge. However, for typical parameters for slit width, disk scale height, scale length, and truncation radius, the effect on the profile shape is insignificant compared to the typical surface brightness fluctuations.

\item $H_3:$ A single exponential profile:
\begin{equation}
f(r) = c_0  e^{-\lambda_1  r} .
\end{equation}
This model describes an unbroken exponential law, in accord with the Freeman Type I profile \citep{freeman}. Its only free parameters are $c_0$ and $\lambda_1$.
\end{itemize}

The second consideration is how we treat the data. We can either examine individual galaxy
profiles in an effort to quantify the range of variation among profiles and whether any trends
become evident, or combine the data into a single spectrum to enable a higher
S/N analysis and, arguably, to reach general conclusions about the galaxy population. We choose
to present the composite analysis first, to define what an ``average" galaxy might look like
and address questions about the profiles from the highest possible S/N spectra. The degree
to which the composite represents a real galaxy rather than an amalgam of disparate objects will
then be discussed when we examine individual galaxy profiles.

\subsection{Composite H$\alpha$ and Continuum Surface Brightness Profiles}
\label{sec_composite}

In this section, we focus on the general properties of the galaxies
in our sample by superposing all individual spectra to create a composite spectrum. We take into account the fact that our sample consists of two subsamples, one of undisturbed, one of warped galaxies. In the case of the warped galaxies, we expect the galaxy isophotes to curve away from the slit position at and beyond the onset of the warp, which is likely to occur around a distance of $R_{25}$ \citep{briggs} and should lead to a steep artificial drop of the surface brightness along the slit. Therefore, we will discuss the profile shapes of the two subsamples separately as well.

In our default analysis, each galaxy is weighted equally, so that the resulting composite profile shape is effectively weighted by surface brightness. Although the average surface brightness scatters within only a factor of $\sim 3$ between most of the galaxies,  there are outliers with very faint or
bright surface brightness values, whose contributions to the composite will be affected by the choice of weighting; in particular, the measured surface brightness integral of IC 2058 is almost three times as high as the second-brightest object. To test whether such outliers dominate our composite profiles and thus bias our conclusions, we carry out an alternative analysis in which the contribution of each individual galaxy is inversely weighted by the integral of its surface brightness profile over radius; in other words, the contributing spectra are normalized to a common mean surface brightness.

\begin{figure}
\includegraphics[width=84mm]{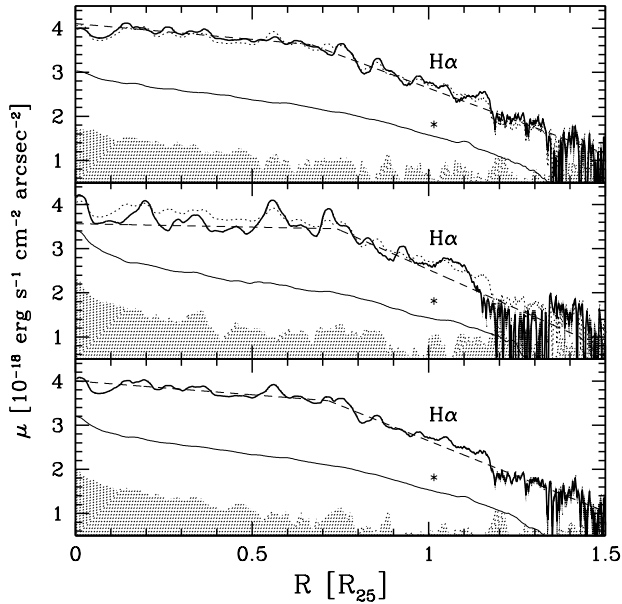}
\caption{H$\alpha$ and stellar continuum composite profile. The panels show, from top to bottom, the undisturbed, warped, and full sample. In each panel, the bold line represents the H$\alpha$ surface brightness, the thin line the stellar continuum (surface brightness per \AA). The dotted line in the upper two panels shows the H$\alpha$ profile from the bottom panel (full sample) to guide the eye. The dashed lines show the best fits to the H$\alpha$ profiles. The shaded region 
indicates the background error in the H$\alpha$ flux. The vertical (flux) scale is linear up to y = 1, which corresponds to $\mu = 10^{-18}$ erg s$^{-1}$ cm$^{-2}$ arcsec$^{-2}$; beyond, each integer step along the abscissa corresponds to a factor of 10.}
\label{fig_ha}
\end{figure}

In Fig. \ref{fig_ha} we present the surface brightness profiles of the H$\alpha$ and stellar continuum emission in our composite spectra. Even cursory visual inspection of the H$\alpha$ and stellar continuum profiles shows that neither a single exponential nor a truncated exponential represent them well. Furthermore, there are subtle differences between the undisturbed and warped composite profiles: The latter are slightly flatter overall and exhibit a particularly strong drop beyond $\sim 1.1 R_{25}$, which may be a signature of the onset of the warp. For these reasons, we decide to fit all composite profiles with broken exponential laws according to Eq. \ref{eq_brokenexp}. It is possible that even a broken exponential law provides only an insufficient representation of the true profile shape, and that additional parameters would be required to model it accurately (e.g., a three-component broken exponential model for the warped objects); however, given the small size of the sample, we decide not to investigate more complex models in this paper.

Given that the constituent galaxies may represent a range of profile shapes,
the uncertainties in these fitting parameters for the composite
profile are likely to be dominated by scatter between the galaxies,
rather than the measurement uncertainties. We therefore determine the
uncertainties on these parameters by bootstrapping, i..e, by randomly
resampling the set of profiles (two profiles per galaxy, representing the two sides) that we superpose to construct the composite, and fitting each of the resulting realizations of the composite spectrum separately. Because of the small size of the sample, we cannot guarantee that this procedure adequately samples the true variance in the parent population, but it provides a representation of the variance within the sample itself.

\begin{figure}
\includegraphics[width=84mm]{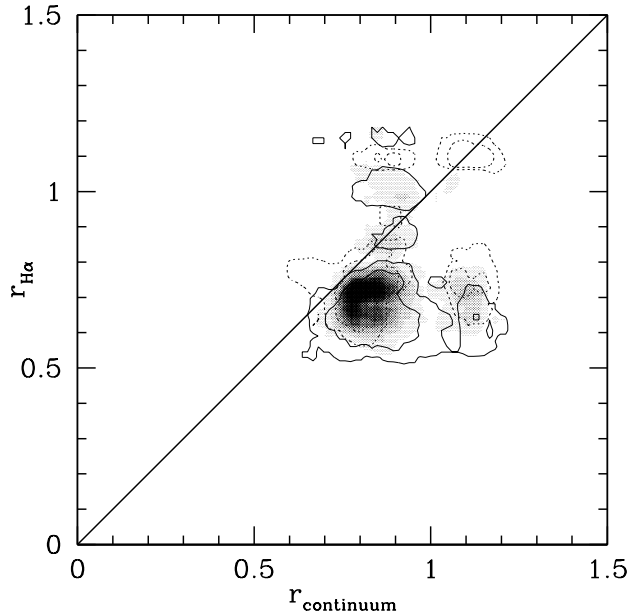}
\caption{Probability distribution of the truncation radii $r_0$ of the H$\alpha$ and stellar continuum components in the composite profile.  Shaded regions show the distribution in the full sample, solid contours in the morphologyicall regular sample, and dotted contours in the warped sample. Contour lines are drawn at the $p=95\%$ and $p=68\%$ contours. Shaded areas represent
parameter combinations with $p > 5$\%, with darker shading representing the more preferred values.}
\label{fig_rr}
\end{figure}

We begin our analysis and comparison of the composite profile shapes by
examining the break radius, $r_0$, for both the H$\alpha$ and stellar components. Fig. \ref{fig_rr} shows the distribution of break radii recovered from our bootstrapping procedure for the undisturbed (solid contours), warped (dashed contours), and full (greyscale) samples. In the full sample, represented by greyscales, the probability distribution for the H$\alpha$ component exhibits a strong peak around $r_0\approx0.7$ and an extended tail towards larger radii, as well as a secondary peak for the stellar continuum break radius at $\sim 1.1 R_{25}$.  The 95\% contours encompass the locus $r_{H\alpha}=r_{continuum}$; therefore, we cannot rule out that the break radii for the stellar continuum and H$\alpha$ components are the same. The undisturbed and warped subsamples exhibit a qualitatively similar behaviour, with the difference that the secondary maximum for an H$\alpha$ break radius around 1.1 $R_{25}$ is much more pronounced in the warped than in the undisturbed or full samples.

\begin{figure}
\includegraphics[width=84mm]{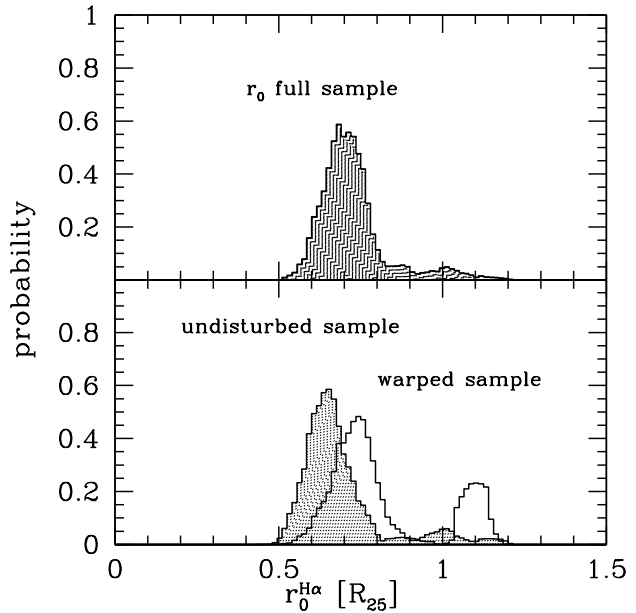}
\caption{Probability distribution for the break radius $r_0$ of the H$\alpha$ surface brightness profile in the full sample (upper panel) and split up into the older sample of galaxies with undisturbed morphology (bottom panel, shaded histogram) and the newer sample of warped galaxies (bottom panel, unshaded histogram). The bimodality of the probability distribution is much stronger in the warped sample; we therefore interpret the peak at $r_0=1.1$ as the onset of the warp. However, both samples independently show evidence for a break in the H$\alpha$ surface brightness profile around $r_0=0.7$.}
\label{fig_r}
\end{figure}

To further explore these results, we show the projection of this probability distribution onto the $r_0^{H\alpha}$ axis in Fig. \ref{fig_r}. All samples exhibit a strong peak in the probability distribution at $r_0\sim 0.7$, as well as a secondary peak around $r_0\sim1.0 - 1.1$. However, this secondary peak is much stronger in the warped than in the undisturbed sample; we therefore believe that it is related to the onset of the warp, where a drop in the surface brightness is expected. Nevertheless, the evidence for the 0.7$R_{25}$ peak is clear even among the warped galaxies, thus justifying the decision to include both undisturbed and warped galaxies in our analysis. These results, modulo slight quantitative differences, are also obtained when weighting the individual spectra inversely by their integrated surface brightness in the composite (not shown in the figure, as the differences are small).

 We conclude 1) that the H$\alpha$ break radius of 0.7 appears across both types of galaxies and so is not a result of warps,  2) that this H$\alpha$ break occurs in unwarped galaxies at $r_0^{H\alpha}=0.65^{+0.05}_{-0.06}$ ($0.68^{+0.08}_{-0.02}$ for the full sample), and 3) that the break radii in the stellar continuum and H$\alpha$ are consistent.

\begin{figure}
\includegraphics[width=84mm]{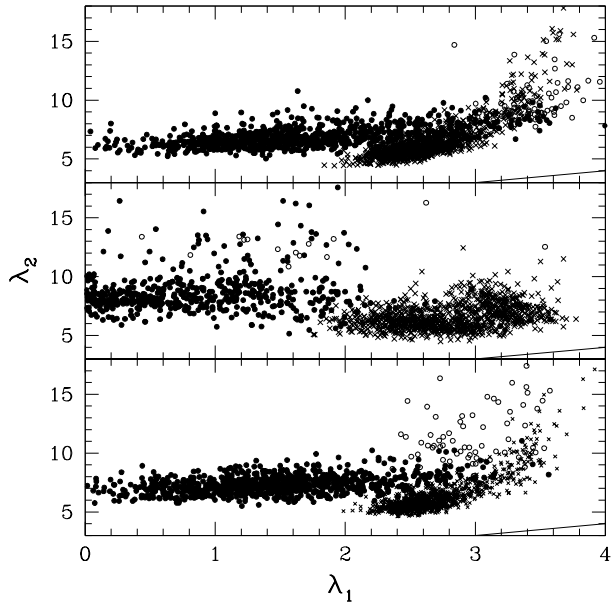}
\caption{Probability distribution for exponential slopes $\lambda_1$
  and $\lambda_2$ in the inner and outer disk, respectively, for the
  H$\alpha$ component (circles) and the stellar continuum
    (crosses). Panels show, from top to bottom, the undisturbed, warped, and full sample. Filled circles mark fits with H$\alpha$   truncation radius $r_0<0.9$, and empty circles fits with $r_0>0.9$. The distribution for the H$\alpha$ profile is bimodal, but the upper peak is associated with the truncation radius $r_0\approx1.0$ seen in the warped galaxy sample.  The truncation is sharper in the H$\alpha$ profile than in the stellar continuum. The line shows the locus $\lambda_1=\lambda_2$, i.e., a single, unbroken exponential profile.}
\label{fig_mm}
\end{figure}

The other important characteristic of the profiles that we consider in our comparison of H$\alpha$ and stellar continuum surface brightness profiles are the inner and outer slopes. Fig. \ref{fig_mm} shows the distribution of inner and outer slopes recovered from the bootstrapping procedure for both components (filled and empty circles for the $H\alpha$, crosses for the stellar continuum). Our discussion will focus on the full sample, but notable differences between the undisturbed and warped samples will be discussed. Our first conclusion from this Figure is that the locus $\lambda_1=\lambda_2$, indicated by a solid line in the bottom right corner of each panel, is inconsistent with both distributions; instead, $\lambda_2>\lambda_1$. This implies that the composite surface brightness profiles decline more steeply in the outer disk than in the inner one, i.e., both the stellar continuum and H$\alpha$ profiles in the composite spectrum are of the Freeman Type II (the sub-exponential case), and hypothesis $H_3$ (Freeman Type I, the exponential profile) is ruled out.

While the stellar continuum fits populate a fairly compact region in
the $\lambda_1-\lambda_2$ parameter space, fits to the H$\alpha$ composite profile exhibit an extended tail. The maximum of the probability distribution for the full sample lies around $\lambda_1\approx1.5$ and $\lambda_2\approx7$, but a significant fraction of realizations are fitted with much steeper slopes both in the inner and outer disk.  Given our previous discussion of a dichotomy resulting from the combination of two samples with different selection criteria, it is natural to ask whether the two subsamples can be identified with distinct loci in Fig. \ref{fig_mm}. Indeed, selecting only those realizations of the bootstrapping procedure that yield best fits at $r_0<0.9$, i.e., those associated with what we consider the true truncation radius, produces fits solely in a well-confined region with $\lambda_1<2.5$ and $\lambda_2<9$, which we have indicated with filled circles. The remaining fits with $r_0>0.9$ are represented by empty circles. We will focus our analysis on the $r_0<0.9$ peak, which we believe to be uninfluenced by the warps in several of our sample galaxies.

A further important observation from Fig. \ref{fig_mm} is that the outer disk slopes $\lambda_2$ for the H$\alpha$ component are slightly displaced from those for the stellar continuum towards steeper slopes, indicating that H$\alpha$ drops off slightly faster. Neither component is extremely large $(>10)$; with a mock catalogue, we have verified that a sharp truncation would yield values of $\lambda_2>20$. Projection effects, as noted earlier, will soften the outer edge of the surface brightness distribution only slightly and do not change this conclusion. Therefore, this result confirms that the truncation in neither component is particularly sudden, i.e., $H_2$ is ruled out for the composite profile, as we had anticipated earlier based on visual inspection alone.

Finally, we turn our attention to the comparison between the H$\alpha$
and stellar continuum profiles. Irrespective of the influence of
warped galaxies on the fit, we note that the two probability
distributions are distinct from each other. In particular, fits to the
stellar continuum are somewhat closer to the condition
$\lambda_1=\lambda_2$, meaning that the truncation is relatively
softer. The fits to the H$\alpha$ component, on the other hand, are
marked by a significantly larger difference between inner and outer
slopes, i.e., the truncation is sharper. If we consider only the peak
associated with the $r_0\approx0.7$ fits, we find the slope
$\lambda_1$ of the H$\alpha$ profile in the inner disk to be fairly
shallow and much shallower than for the stellar continuum.
In the outer disk, however, the situation appears reversed: The
  values of $\lambda_2$ appear to indicate a slightly steeper slope
  for the H$\alpha$ component than for the stellar continuum. 
  
  The preceding observations also apply when we consider the samples of warped and unwarped galaxies separately; most differences are quantitative, but not qualitative. The most striking difference is the fact that the surface brightness profile of the warped galaxies is fitted with shallower inner slopes. One possible explanation is that the warped galaxies were selected to be highly edge-on systems, so that the surface brightness measured in the inner disk is strongly affected by dust in the galactic plane. However, since our analysis is not designed to probe the inner disks, we cannot pursue this observation further. More relevant to our investigation, in the undisturbed subsample, the tail of the probability distribution of fits to the stellar continuum profile extends to much steeper inner and outer slopes. This leads to a significant overlap between the probability distributions for the H$\alpha$ and stellar continuum component. However, further investigation reveals that this tail is associated with fits with a very large break radius ($>R_{25}$); the apparent overlap in the fit parameters for the H$\alpha$ (which all have a much smaller break radius) and stellar continuum profiles is therefore caused only by the chosen projection, and the conclusion above that the two probability distributions are distinct from each other still holds.

\begin{figure}
\includegraphics[width=84mm]{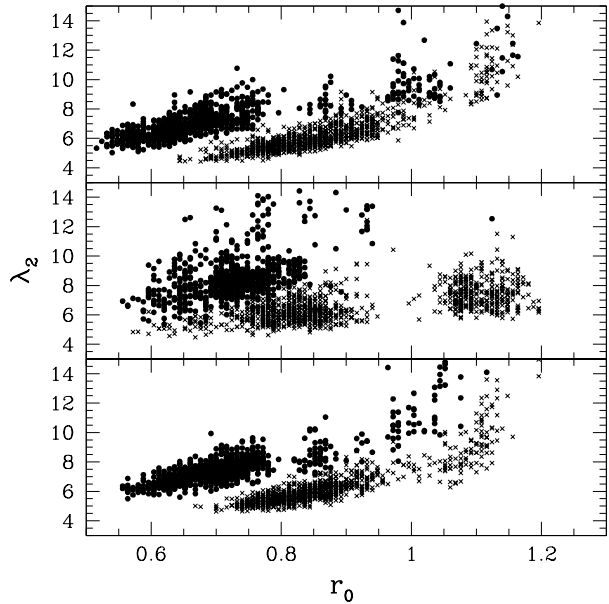}
\caption{Probability distribution for the outer-disk exponential slope
  $\lambda_2$ vs. break radius $r_0$ for the H$\alpha$ (dots) and
  stellar continuum surface brightness (crosses) profiles. Panels show, from top to bottom, the undisturbed, warped, and full sample. The two
  profile shapes are inconsistent with each other; for a given break
  radius $r_0$, the H$\alpha$ component drops more steeply in the
  outer disk.}
\label{fig_mr}
\end{figure}

  To better decide whether the stellar continuum and H$\alpha$ components follow significantly different spatial distributions or not, we examine an additional projection of the distribution of best-fit
  parameters. In Fig. \ref{fig_mr}, we examine the projection of the profile fit parameters onto the $r_0$-$\lambda_2$ plane, i.e., we plot the break radius versus the outer-disk profile slope. Again, filled dots represent realizations of the H$\alpha$
  composite profile, and crosses realizations of the stellar continuum
  profile. As previously, in order from top to bottom, the panels show the undisturbed, warped, and full samples. The fits to the H$\alpha$ profile at $r_0\approx1.1 R_{25}$ in the warped sample occur at very steep values of $\lambda_2>15$ and are thus outside the scale of this plot.  In this figure, we clearly see that the overlap between the best-fit parameters for H$\alpha$ and stellar continuum is
  negligible for all three samples: For a given break radius $r_0$, the H$\alpha$ profile does not trace the stellar continuum, but declines significantly more steeply (but still with a finite slope).

\subsection{Individual Galaxies}
\label{sec_individual}

\subsubsection{Statistical tests}

Throughout this section, we will test the likelihood of a given hypothesis describing the shape of the H$\alpha$ or stellar continuum surface brightness profiles. These tests are based on $\chi^{2}$ quantifiers. However, it is not possible to quantify the absolute goodness of a fit using the $\chi^{2}$ estimator alone, because the scatter of the surface brightness measurements around the fit is dominated not by statistical measurement errors, but by intrinsic surface brightness fluctuations of the emission along the slit. These surface brightness fluctuations are not reflected in our error bars; therefore, the formal quality of a $\chi^{2}$ fit is always poor. Instead, we use a ratio 
\begin{equation}
r_{\chi^{2}}=\frac{\chi^{2}(H_n)}{\chi^{2}(H_1)}
\end{equation}
as our statistical quantifier, where $\chi^{2}(H_n)$ is the $\chi^{2}$ value for a given hypothesis for the profile shape. The hypothesis $H_1$ is our most flexible one - it describes a broken exponential profile with a normalization factor $c_0$, a break radius $r_0$, and slopes $\lambda_1$ for $r<r_0$ and $\lambda_2$ for $r\geq r_0$. All other hypotheses $H_n$ that we test are subsets of these four-parameter profiles, obtained by constraining one or more of the parameters to specific value. In general, the four-parameter model yields the best fit due to its having the largest number of degrees of freedom. This applies even in cases where $H_n$ does provide a correct description of the underlying surface brightness profile because small deviations from this shape due to noise and fluctuations are generally fitted better by the four-parameter hypothesis $H_1$. However, a very large ratio $\frac{\chi^{2}(H_n)}{\chi^{2}(H_1)}$ indicates that the fit under hypothesis $H_n$ is significantly worse than achievable with a four-parameter model, and that $H_n$ has to be rejected.  The virtue of this approach is that it is insensitive to the assumed uncertainties in the surface brightness measurement; the normalizations of the uncertainties cancel out as we form a ratio of $\chi^{2}$ values. However, a complication lies in finding the critical threshold value for the $\chi^{2}$ ratio beyond which we can rule out that the improvement in the quality of the fit results simply from the ability to fit random fluctuations better with more free parameters, and is indeed indicative of a true preference of $H_1$ over $H_n$.

To calibrate  $r_{\chi^{2}}$  as a statistical measure
of significance, we use a Monte Carlo simulation that generates mock
surface brightness profiles conforming to the tested hypothesis
$H_n$. The mock profile is created as a superposition of individual surface brightness fluctuations. The spatial shape of each profile is described by a Gaussian function with a fixed width. The number of fluctuations that are added up in each row of the simulated profile are determined from an assumed average amplitude and Poisson statistics. The actual flux of each Gaussian fluctuation is drawn from a Gaussian probability distribution over logarithmic flux with a pre-defined width. In addition, we add fluctuations with a fixed, small amplitude to simulate fluctuations in the background. The average amplitude, spatial width, width of the flux probability distribution and background fluctuation amplitude are the parameters of this model; in addition, we have two parameters controlling the size of the simulated uncertainties. We have run a series of simulations to find the parameters that recreate profile shapes most similar in appearance to the observed ones by fitting broken-exponential models to each mock profile and then using three statistics to compare them to the observations: 1) the ratio of the integrated residuals between 0.2 and 1.5 R$_{25}$ to the total integrated profile in that
range. 2) the cumulative residual  between 0.2R$_{25}$ and
an outer radius $x$, normalized by the total integrated residual
between 0.2 R$_{25}$ and 1.5R$_{25}$. 3) the cumulative $\chi^{2}$ between 0.2R$_{25}$ and
an outer radius $x$, normalized by the total integrated $\chi^{2}$
between 0.2 R$_{25}$ and 1.5R$_{25}$.
The first of these statistics characterizes the magnitude of the fluctuations relative to the
total flux. The second characterizes the radial behavior of the
size of fluctuations. The third characterizes the relative goodness-of-fit
as a function of radius. Of particular importance is that, both in the mocks and in
the real profiles, contributions to $\chi^{2}$ come primarily from
intermediate radii around $r\approx 0.7$, i.e., $\chi^{2}$ is
particularly sensitive to the region where we locate the break in the
composite profile.

We determine the $\chi^{2}$ ratio, $r_{\chi^{2}}$,
for a given hypothesis $H_n$, and compare it to the scatter of
$r_{\chi^{2}}$ in our mock profiles. If $r_{\chi^{2}}$ is outside the
range reproduced by 95\% of the Monte Carlo realizations, we reject
the hypothesis with 95\% confidence.

\label{sec_statistics}

\subsubsection{Individual galaxy H$\alpha$ surface brightness profiles}

For our analysis of the H$\alpha$ surface brightness profiles in the individual galaxies, we estimate the radial range of H$\alpha$ detections in each galaxy. Although our fitting procedure includes the entire radial range between $0.2 < R/R_{25} < 1.5$, the position of the outermost detection additionally provides a lower limit on the position of the truncation in H$_2$.

To determine the radius of the outermost significant H$\alpha$ detection, we integrate the total H$\alpha$ flux between a given radius $R$ and the largest radius considered in our analysis, $1.5 R_{25}$. To determine the limiting radius $R_{lim}$, we then use two alternative definitions. The first, $R_{lim}^{a}$, is the radius of the outermost H$\alpha$ detection with $> 3\sigma$ significance between $R$ and $1.5 R_{25}$. The second, $R_{lim}^{b}$, is the smallest radius $R$ for which the 
H$\alpha$ flux is $< 2\sigma$ significant. The first definition may include isolated H$\alpha$ regions at large radial distance from the galaxy center, while the second definition measures only the extent of contiguous H$\alpha$ emission and does not include isolated sources in the outer disk. However, in practice, both definitions yield very similar results, except in the case of ESO 478- G011, where an H$\alpha$ detection of low significance pushes the outermost radius to $R_{lim}\approx1.45$, while contiguous H$\alpha$ emission is only detected out to $R_{lim}\approx1.34$. We list $R_{lim}$ for all galaxies in our sample in Table \ref{tab_outermost}.  

We have not found obvious correlations between the extent of the H$\alpha$ emission and other global galaxy properties, such as the total H$\alpha$ emission, apart from a very weak correlation between $R_{lim}$ and the integrated H$\alpha$ flux beyond $r_0$, which indicates that only galaxies with large $R_{lim}$ may have significant flux in this second, outer-disk exponential component.

\begin{figure}
\includegraphics[width=84mm]{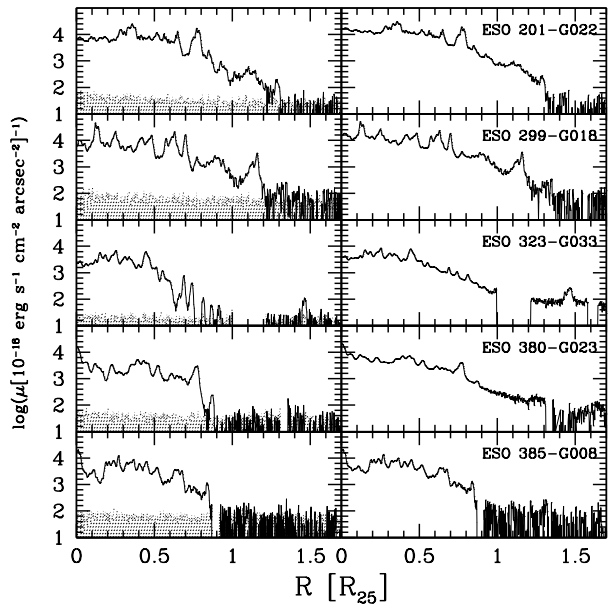}
\caption{H$\alpha$ (left) and stellar continuum (right) surface brightness profiles for individual
  galaxies, here from the morphologically undisturbed sample. Shaded regions in the H$\alpha$ profile indicate background uncertainties.}
\label{fig_profiles1}
\end{figure}

\setcounter{figure}{5}
\begin{figure}
\includegraphics[width=84mm]{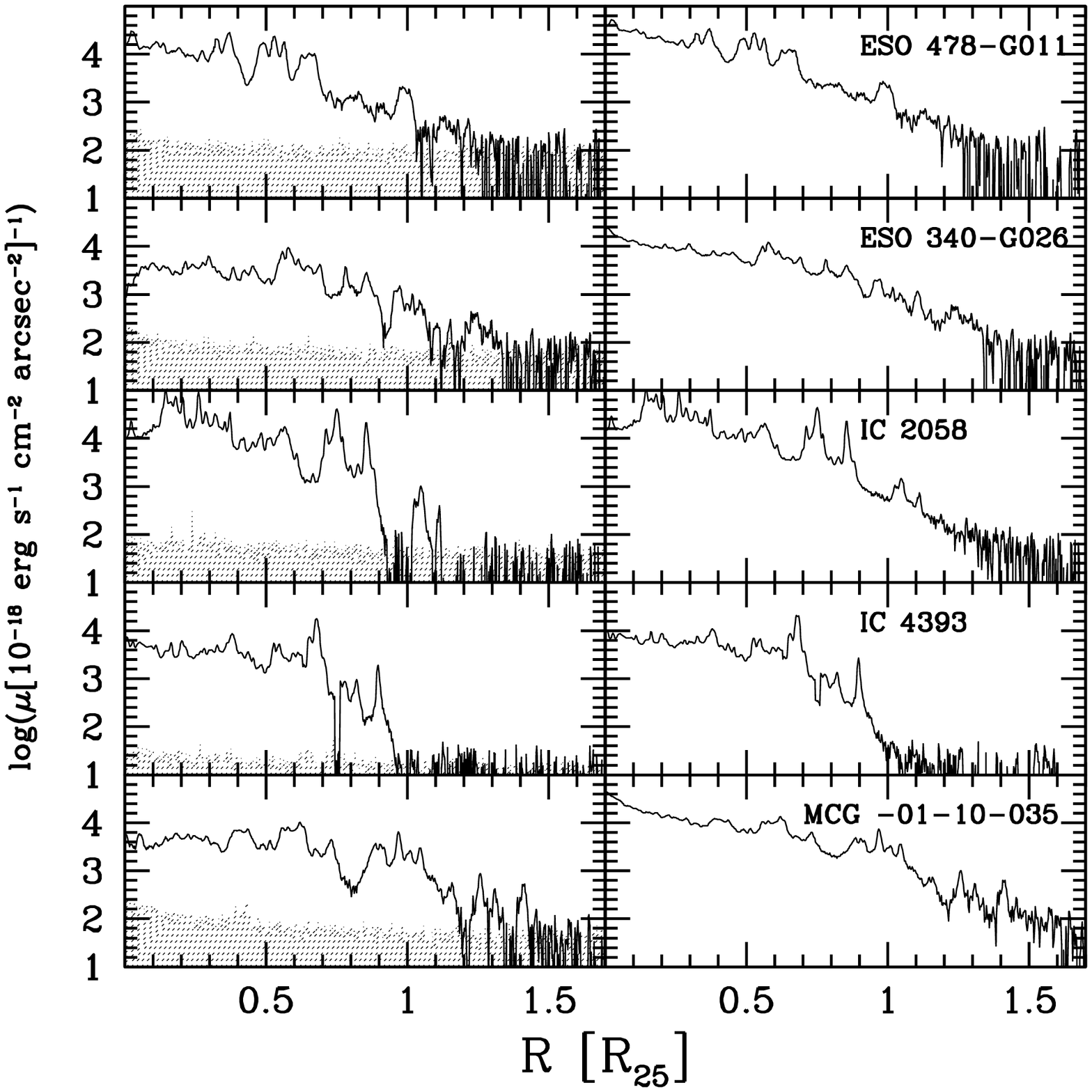}
\caption{(continued)}
\label{fig_profiles2}
\end{figure}

\begin{figure}
\includegraphics[width=84mm]{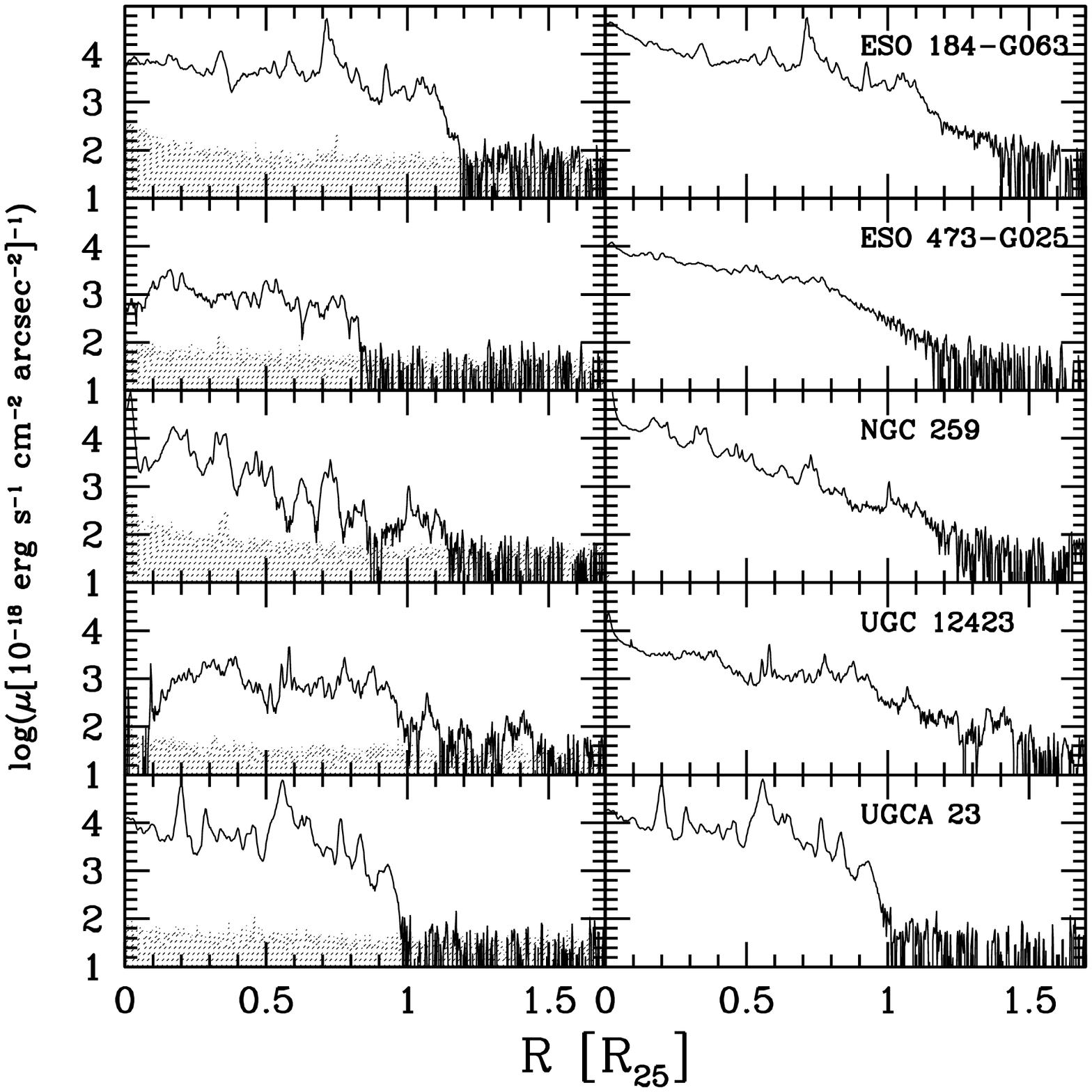}
\caption{H$\alpha$ (left) and stellar continuum (right) surface brightness profiles for individual
  galaxies, here from the warped sample. Shaded regions in the H$\alpha$ profile indicate background uncertainties.}
\label{fig_profiles3}
\end{figure}

The H$\alpha$ surface brightness profiles of the individual galaxies are shown in Fig. \ref{fig_profiles1} (which consists of two parts) for the morphologically regular sample, and in Fig. \ref{fig_profiles3} for the warped sample.
 By visual inspection, we identify a break in several of these galaxies, while others appear consistent with single exponentials.  However, in general, there are large surface brightness fluctuations that preclude us
reaching specific conclusions from a cursory viewing.

For this reason, a careful statistical analysis is required. We examine whether the individual galaxies are consistent with specific hypotheses for the shape of the composite profile in the manner described in \S \ref{sec_statistics}: We calculate the $\chi^{2}$ ratio between fits with the specific hypothesis and a more general four-parameter, broken exponential fit, and then determine, using a Monte Carlo simulation, if this ratio indicates that a four-parameter model provides a significantly better fit and that the tested hypothesis must be discarded. In all cases, we construct the input profile shape for the Monte Carlo simulation on the basis of the best-fit parameters for the composite profile under the given hypothesis (broken, truncated, single-exponential). We find no qualitative global differences when using the two different weighting schemes in our construction of the composite spectra, although a few individual galaxies shift from being consistent with a given hypothesis to being inconsistent and vice versa. For the remainder of this discussion, we use the composite constructed from the unweighted combination of individual spectra.

Using the lessons learned from the composite spectrum and the details evident in the individual spectra, we modify our set of models to include a version of $H_1$, the broken exponential, that sets the break radius to 0.7 (model $H_{1a}$), a version of $H_2$, the truncated exponential, that sets the radius of truncation to be that of the outermost H$\alpha$ detection (model $H_{2a}$), a single-exponential model ($H_3$), and a new model that uses the best-fit composite profile (model $H_4$ with free parameter $c_0$ to allow renormalization). The fit quality is evaluated only between 0.2 and 1.5 $R_{25}$.

We present the results of this analysis in Table \ref{tab_indresults}. The entries in the Table give the fraction of realizations for which the ratio $\chi^{2}_{Hn}/\chi^{2}_{H1}$ exceeds the value obtained from the observed profile. Therefore, low values indicate that hypothesis $H_1$ provides a significantly better fit to the surface brightness profile than the alternative. We highlight entries $<0.05$, i.e., those that formally indicate that the considered hypothesis is inconsistent with the observed profile shape at the 2$\sigma$ level. 

\begin{table}
 %\centering
 %\begin{minipage}{140mm}
  \begin{tabular}{@{}lrrrrrr}
  \hline
Name  & $H_2$ & $H_3$ & $H_{1a}$ & $H_4$ & $H_{2a}$ & Warp? \\
 \hline
   ESO 201- G 022 & {\bf 0.00}  & {\bf 0.00}  &       0.06  & {\bf 0.01}  & {\bf 0.00}	& N \\
  ESO 299- G 018 & {\bf 0.04}  &       0.69  &       0.91  &       0.19  & {\bf 0.02} & N \\
  ESO 323- G 033  & {\bf 0.00}  & {\bf 0.00}  & {\bf 0.00}  & {\bf 0.00}  & {\bf 0.00} & N \\
  ESO 380- G 023  & {\bf 0.02}  & {\bf 0.00}  &       0.44  & {\bf 0.00}  & {\bf 0.00} & N \\
  ESO 385- G 008  & {\bf 0.03}  &       0.08  &       0.82  & {\bf 0.00}  & {\bf 0.00} & N \\
    ESO 478- G 011  &       0.06  &       0.92  &       0.90  & {\bf 0.05}  & {\bf 0.04} & N\\ 
      ESO 340- G 026  & {\bf 0.00}  & {\bf 0.03}  &       0.80  &       0.65  & {\bf 0.00}   &N\\
        IC 2058  &       0.07  &       0.37  &       0.67  & {\bf 0.00}  & {\bf 0.00} & N\\ 
  IC 4393  & {\bf 0.02}  & {\bf 0.01}  &       0.06  & {\bf 0.00}  & {\bf 0.00} & N\\
    MCG -01-10-035 & {\bf 0.02}  &       0.68  &       0.77  &       0.12  & {\bf 0.02} & N\\
  ESO 184- G 063  & {\bf 0.02}  & {\bf 0.00}  &       0.39  & {\bf 0.00}  & {\bf 0.00}	& Y\\
  ESO 473- G 025 & {\bf 0.03}  & {\bf 0.00}  &       0.24  & {\bf 0.01}  & {\bf 0.01} &Y\\
  NGC 259 & {\bf 0.03}  &       0.86  &       0.89  & {\bf 0.01}  & {\bf 0.03} & Y \\
  UGC 12423 & {\bf 0.01}  &       0.21  &       0.47  & {\bf 0.04}  & {\bf 0.00} & Y \\
  UGCA 23 & {\bf 0.00}  & {\bf 0.00}  &       0.89  &       0.11  & {\bf 0.00} & Y\\
\hline
\end{tabular}\\
\caption{Fraction of simulations that exceed the observed $\chi^{2}_{Hn}/\chi^{2}_{H1}$ ratio for a given galaxy, where $H_n$ is one of the following: $H_2$: trunated exponential, $H_3$: single exponential, $H_{1a}$: broken exponential with $r_0=0.7$, $H_4$: scaled version of composite best fit, $H_{2a}$: truncated exponential extending to last significant H$\alpha$ detection.}
\label{tab_indresults}
\end{table}

We now discuss the results in terms of specific questions.

\subsubsection{Are the individual galaxies consistent with
  a truncated exponential surface brightness profiles?}

Hypothesis $H_2$ is that of an exponential profile with a sharp truncation. Most galaxies in our sample, whether warped or not, are inconsistent with the truncated profile shape. The two cases that are marginally consistent with a sharp truncation are ESO 478-G011 and IC 2058. Visual inspection supports this for IC 2058; in the case of ESO 478-G011, the best-fit truncation radius is at the very large radius of $1.34 R_{25}$, so that the truncated profile is essentially indistinguishable from a single exponential shape.
 
We now restrict this hypothesis. In many of our sample galaxies, H$\alpha$ is observed to large radii, as Table \ref{tab_outermost} shows. Such detections are formally inconsistent with an exponential profile truncated at smaller radii. Therefore, we impose the additional constraint that the truncation radius be located just beyond the radius of the outermost H$\alpha$ detection and choose $R_{lim}^b$ for this purpose. In Table \ref{tab_indresults}, the results of this analysis are shown in column $H_{2a}$. None of the galaxies in our sample are described well by a profile truncated at the last significant H$\alpha$ detection.

Lastly, perhaps a sharply truncated profile is ruled out by this analysis only because the truncation is softened in projection due to the nearly-edge-on configuration. Simple simulations of the projected profiles that would be obtained from such a configuration show that, even for an almost perfectly edge-on geometry, the softening effect towards the edge of the H$\alpha$ disk would be fairly subtle and unlikely to exceed the surface brightness fluctuations observed in the real spectra. Therefore, this analysis confirms that a broken-exponential profile with a finite outer-disk slope provides a significantly better fit to the individual galaxy profiles than a sharply truncated single exponential law.

\subsubsection{Are the individual galaxies consistent with an unbroken exponential profile?}

 Next, we consider $H_3$, the hypothesis that the H$\alpha$ surface brightness profile is described by a single, unbroken exponential. This model corresponds to the Freeman Type I profile, which is observed to hold for the stellar continuum of some galaxies \citep{blandhawthorn2005,pohlen2,erwin} out to extremely large radii.
 
Our sample is dominated by galaxies with broken exponential profiles in the stellar continuum (Figures \ref{fig_profiles1} to \ref{fig_profiles3}), where visual inspection identifies only few objects as being consistent with a single exponential profile. For about half of the H$\alpha$ profiles in our sample, whether they represent normal or warped galaxies, we also rule out the
unbroken exponential profile shape ($H_3$), as Table \ref{tab_indresults} shows.
The exceptions among the unwarped galaxies may be of particular interest because the existence of a critical threshold surface gas density beyond which the specific star formation rate in a galaxy changes abruptly or even drops to zero has been suggested repeatedly in the literature \citep{martinkennicutt}. The objects for which we cannot conclusively rule out an unbroken exponential law are ESO 299-G018, ESO 385-G008, ESO 478-G011, IC2058, MCG -01-10-035, NGC 259, and UGC 12423. In some cases, this can be attributed to the strong fluctuations, which lower the quality of even the broken-exponential fit and reduce our ability to discriminate between different hypotheses. Visual inspection does single out two of these objects, ESO 478-G011 and NGC 259, as the best candidates for an unbroken exponential profile; these are also the two objects best fitted under $H_3$. This is surprising for NGC 259, which is listed as a warped galaxy; however, within our warped sample, this object is the least-inclined, and the classification as a warped object may be due to a prominent spiral arm, rather than to an actual vertical bending of the disk, which may explain why we can trace a regular, possibly unbroken H$\alpha$ profile out to $\sim 1.35\times R_{25}$.

We conclude that, just like the composite profile, the individual galaxy H$\alpha$ profiles are generally inconsistent with unbroken exponentials out to large radii, but that a significant number of exceptions might exist in our sample.

\subsubsection{Are the individual galaxies consistent with a universal break radius?}

The third question we address is whether the individual
galaxies might all share a common break
radius, $r_0\approx0.7$, as measured in 
the composite profile. Is this radius a universal property of late-type
galaxies, or simply an average for a population with an
intrinsically large scatter? We address this question by starting with 
H$_1$, and then fixing the
break radius at $r_0=0.7$, but leaving $c_0$, $\lambda_1$, and
$\lambda_2$ as free parameters. We refer to this model as H$_{1a}$. 
The result, presented in the $H_{1a}$ column of Table \ref{tab_indresults}, shows that all but one
(ESO 323-G033) of our galaxy profiles are consistent ($< 2\sigma$ discrepant) with broken exponential profiles with a break radius of $r_0=0.7$. ESO 323-G033 bears the distinction of being consistent with neither of the four alternative profile shapes we are considering here. Because a single inconsistency at the 2$\sigma$ level in a sample of 15 is within statistical expectations, we conclude that the hypothesis of a ``universal" break radius for late-type galaxies like we have selected is consistent with these data.

\subsubsection{Are the individual galaxies consistent with the composite profile?}

Extending the previous analysis, we now examine whether the
composite profile calculated in \S \ref{sec_composite} is itself an acceptable representation of all individual galaxies in the sample. Again, we create mock profiles, this time with the composite
spectrum for which we allow a renormalization.
The results (Table \ref{tab_indresults}) show that the H$\alpha$ profiles of most of our galaxies are inconsistent with the composite; there are only four exceptions: one from the undisturbed sample (ESO 299-G018) and three from the warped sample (ESO 340-G026, MCG -01-10-035, UGCA 23). We conclude that there are real variations among the galaxies and that
the composite spectrum is not necessarily a good representation of a galaxy chosen at random, with the exception of the break radius at $r_0=0.7 R_{25}$.

\section{Discussion}

We find that more than half of our galaxies do not have a Freeman Type I H$\alpha$ surface brightness profile (a single, unbroken exponential). Visual inspection demonstrates that the profiles are rather of the downbending (sub-exponential) Type II class, and are thus qualitatively following the behaviour of the stellar continuum, i.e., their surface brightness profiles may be described as broken exponential profiles
with a characteristic break radius and a steeper slope in the outer
disk than in the inner disk. The H$\alpha$ break radius at $r_0\approx 0.7$ R$_{25}$ appears surprisingly well-constrained and is observable both in the unperturbed and the warped subsamples.

Observations of a break in the H$\alpha$ surface brightness profile 
are not new. Breaks or truncations in the H$\alpha$ profile have been noted for many years \citep[cf.][]{martinkennicutt}, and a star formation threshold, i.e., a critical gas surface sensitivity below which star formation cannot proceed or proceed only very inefficiently has often been suggested as the most likely explanation. Other breaks or truncations are expected at very large radii, where ionization by the cosmic ultraviolet background may become a significant contributor to outer-disk H$\alpha$ emission and where the baryonic disk ends;
however, both of these effects are likely occur at or beyond 1.5 $R_{25}$, which is a typical outer
radius of HI disks, and at surface brightness levels fainter than what we can reliably probe with these data. What is new here is that we quantify the radial behavior of the emission beyond this
break radius and out to large radii, $\sim 1.5R_{25}$. 

It is tempting to identify the clear break in H$\alpha$ with the H$\alpha$, and possibly star formation, threshold well-known from the literature \citep{martinkennicutt}, but if so, then this ``threshold" is clearly not the end of star
formation because H$\alpha$ emission continues with some regularity beyond.
Furthermore, if the threshold position depends on local physical conditions,
it is remarkable how well-defined the location of $r_0$ is with respect to $R_{25}$.

It is also tempting to identify the break radius in the stellar continuum that we observe
with that observed by \citet{pohlen}, and then
further associate it with our observed H$\alpha$ break radius because
the two are consistent within 2$\sigma$ in our data. However, the outer-disk
exponential slopes are different for the two components, with the 
stellar continuum being more extended, suggesting that there is
not a 1:1 correspondence between the surface brightness profiles of
the two components. The physics determining the two are therefore more complex
than a simple relationship between star formation and stellar populations.

If H$\alpha$ is a linear tracer of star formation, the current stellar continuum surface brightness distribution cannot have been produced by in-situ star formation, unless star formation was 
more extended in the past. Because the 
currently favored paradigm is that galaxies form ``inside-out" \citep{trujillo,williams}, we
consider it more likely that either the outer-disk stellar population has resulted from a redistribution of stellar mass (for example, as a result of tidal interactions with other galaxies \citep{stelios} or secular dynamical evolution \citep{roskar}), or that H$\alpha$ is an increasingly biased tracer of star formation towards large radii. 

No less important than the observation of a break in the H$\alpha$ surface brightness profile between the inner and the outer disk is the complementary observation that H$\alpha$ emission is not sharply truncated. We observe H$\alpha$ emission as far out as $1.5 R_{25}$, and find its surface brightness profile to continue with some regularity, possibly even following an exponential law, but with a steeper slope. The observation of such extended emission is consistent with observations in recent years of extended UV emission around several galaxies \citep{thilker,zaritskychristlein}. It demonstrates that star formation can occur even in the extreme outer-disk regime, where gas surface densities are presumably very low. At the same time, the presence of the break in the H$\alpha$ profile suggests that, even if star formation is not completely suppressed in the outer disk, it appears to proceed under different physical conditions. 

The finding that the outer disk UV emission, which is also an
indicator of recent star formation, does not exhibit a break \citep{boissier,boissier2}, has given rise to the suggestion that
the star formation rate profile is also unbroken. A possible explanation of the divergent behaviour 
between the H$\alpha$ and UV profiles is that the truncation of H$\alpha$ emission 
is a stochastic effect due to
the number of star formation regions with sufficiently massive, ionizing stars
dropping below unity \citep{boissier}. However, in our
observations, which reach much higher sensitivies than narrow-band imaging
studies and superpose multiple galaxies, we detect emission at extremely high levels of confidence,
at and beyond the hypothetical truncation radius. Therefore, there are numerous ionizing
sources in the outer disk. Nonetheless, we see a clearly-defined break in the H$\alpha$ surface brightness profile both in individual galaxies and in the composite; we conclude that the break in the H$\alpha$ profile is real, but that it does not mark the end of star formation or the end of the formation of sufficiently massive stars.

 The apparent disagreement with the UV surface brightness profile is thus puzzling. If we interpret the UV observations to mean
that star formation is not subject to a sudden break, then H$\alpha$ must be a biased tracer of star formation and suppressed at large radii. The ratio of UV to H$\alpha$ emission must change continuously with radius. One potential cause of such an effect is an increasingly bottom-heavy initial mass function with radius, either as a result of a real change in the functional form or from statistical effects related to the increasingly poorer sampling of the upper stellar masses as total mass decreases. A scenario in which the {\it integrated galactic} IMF (which is probed here) may vary with environment as a result of variations of the mass function of molecular clouds across the galaxy, while the shape of the IMF itself would be preserved, has been presented by \citet{kroupa}. 

Another potential cause is  a transition from ionization-bounded to density-bounded HII regions. 
If a larger fraction of ionizing photons escape without being converted to H$\alpha$ at larger radii
the UV to H$\alpha$ ratio would change. Within inner galaxy disks the effect seems to go the other way,  with high-luminosity star-forming regions, rather than low-luminosity ones, appearing to be density bounded \citep{beckman}, but the situation in the outer disk could be quite different.

A final explanation for the increasing bias between stellar continuum and H$\alpha$ emission in the outer disk is that outer-disk star formation may be an intermittent phenomenon, and short, but correlated, episodes of star formation in the past may have created the present extended stellar population before returning to the low-level state that we observe in our sample.
 A similar scenario, involving bursts of star formation across the entire disk, has been proposed by \citet{boissier3} based on GALEX observations of UV colors of outer galaxy disks. However, this scenario does not appear to readily explain why the UV surface brightness profile would not exhibit a truncation, unless the galaxy was observed shortly after such a burst, so that the UV outer disk would still be dominated by stars formed during the recent star formation episode, while H$\alpha$ emission, which traces star formation only over very short timescales, has already abated.

\citet{bakos} find that the stellar surface mass density
also exhibits a much less pronounced break than the optical light of
the stellar continuum. They suggest that a change in the composition of
the stellar population, rather than an actual break in the mass
distribution, is responsible for the observed break in the optical
continuum surface brightness. In this
context, it is surprising that both the stellar surface mass density
and the integrated, recent star formation rate (as measured by UV emission) supposedly
show no or only weak breaks, but both stellar
continuum and H$\alpha$ do. 

As exceptions to the rule, those galaxies where we find no evidence
for a truncation or break in the H$\alpha$ emission deserve particular
attention, as any theory attempting to explain the presence of a break
in the H$\alpha$ profile --- for example, by invoking a critical gas surface
density threshold below which star formation is inefficient --- must
also be able to provide an explanation for objects that apparently
violate the rule. Of the objects that are formally consistent
with an unbroken exponential profile, NGC 259 and ESO 478-G011 may be interesting cases
for follow-up studies.

A final consideration must be given to the effect of dust on the H$\alpha$ surface brightness profiles, and, in particular, to the question whether dust absorption could influence our measurement of the characteristic break radius and possibly even our classification of the composite H$\alpha$ profile as a (sub-exponential) Type II, rather than a (single exponential) Type I. If dust extinction sets in around this radius and increases continuously towards the galaxy center, absorbing an increasing fraction of H$\alpha$ flux, it may artificially flatten the profile slope and possibly even create the false impression of a broken exponential shape. This scenario, however would imply that the unabsorbed H$\alpha$ profile could be recreated by extrapolating the outer-disk H$\alpha$ profile back towards the center, implying that the real H$\alpha$ surface brightness in the inner regions would be more than an order of magnitude higher than observed here. As dust shielding is unlikely to be uniform (even in cases of nearly perfect edge-on alignment, the slit intercepts regions both on and off the central dust lane), at least some sight lines should permit glimpses of these substantially higher surface brightness levels, leading to a larger dynamic range in the fluctuations of the inner-disk surface brightness profile. This is not the case; even the highest H$\alpha$ peaks remain far below the levels expected from interpolating the outer-disk profile, confirming that the break is indeed real. Dust shielding may, however, plausibly modify the measured value of the inner-disk profile slope by smaller amounts.

An additional argument against dust obscuring a large fraction of the H$\alpha$ flux comes from the rotation curves. If dust played a significant role up to the break radius of 0.7 $R_{25}$, contributions to the observed H$\alpha$ flux would be weighted towards gas on the outskirts of the galaxy along the edge closer to the observer. 
We would therefore expect the rotation curve to rise very slowly in the inner disk and not flatten before it reaches this point. However, most rotation curves in this sample are already flat or close to flat well before reaching 0.7 $R_{25}$ \citep{christleinzaritsky,christleinbh}, indicating that the H$\alpha$ emission at this point already samples the kinematics of the gas along the entire sight line, and that the observed kinematics and flux cannot be strongly biased by internal dust extinction.

\section{Conclusions}

We have presented a study of H$\alpha$ emission and stellar continuum in 15 low-redshift, late-type, edge-on galaxies. The use of deep long-slit spectroscopy, rather than narrow-band imaging, allows us to probe H$\alpha$ emission to very faint levels, $\sim$10$^{-18}$ erg s$^{-1}$ cm$^{-2}$ arcsec$^{-2}$. H$\alpha$ emission is traced out to 50\% beyond the R$_{25}$ radius, roughly twice as far as typical H$\alpha$ rotation curves in the literature. It is thus possible to probe the outer galactic disks, which have hitherto primarily been a domain of radio astronomy, in H$\alpha$ emission, and thus simultaneously gain information about kinematics, star formation, metallicity, and stellar continuum with arcsecond-scale spatial resolution.

In the present paper, we have focused on the properties of the surface brightness profile of H$\alpha$ emission and compared them to those of the stellar continuum. Our conclusions are:
\begin{itemize}
\item In this sample, which is dominated by Freeman Type II (``sub-exponential") profiles in the stellar continuum (a broken exponential profile with a steeper slope in the outer than the inner disk), H$\alpha$ emission in the composite profile also follows a Type II profile. For the H$\alpha$ composite profile, we can rule out both an unbroken single exponential profile as well as a truncated profile, for which the flux drops to zero at a given break radius.
\item There is a well-defined break radius in the composite H$\alpha$ surface brightness profile at $r_0\approx0.7$ R$_{25}$. 
\item The H$\alpha$ and stellar continuum distributions may be consistent with a single break radius $r_0$, but the stellar continuum profile drops more slowly in the outer disk.
\item The presence of a clearly defined break in the H$\alpha$ surface brightness profile, despite the fact that our observations of the outer disk slope are clearly not limited by Poisson noise, indicates that any apparent deficit of H$\alpha$ emission in the outer disk is not, in general, a stochastic effect arising from low number statistics. The break in H$\alpha$ is real, but it is not a truncation, as evidenced by 
the common presence of low-level outer-disk H$\alpha$ emission to $\sim 1.5 r_0$. 
\item For most objects in our sample, a truncated exponential as well as an unbroken exponential can be ruled out even on an individual basis, thereby requiring a broken exponential profile. All but one of our galaxies are individually consistent with a broken exponential profile, with a fixed H$\alpha$ break radius at $r_0=0.7 R_{25}$, showing that the conclusions drawn from the composite profile describe not just a meaningless average, but are representative of at least a significant fraction of individual galaxies.
\end{itemize}

\section*{Acknowledgments}

Based on observations obtained at the Gemini Observatory (program ID GS-2005A-C-4), which is operated by the
Association of Universities for Research in Astronomy, Inc., under a cooperative agreement
with the NSF on behalf of the Gemini partnership: the National Science Foundation (United
States), the Particle Physics and Astronomy Research Council (United Kingdom), the
National Research Council (Canada), CONICYT (Chile), the Australian Research Council
(Australia), CNPq (Brazil) and CONICET (Argentina).

 Based on observations made with ESO Telescopes at the Paranal
 Observatories under programme ID $<$074.B-0461$>$ and$<$079.B-0426$>$ . 

This research has made use of the NASA/IPAC Extragalactic Database (NED) which is operated by the Jet Propulsion Laboratory, California Institute of Technology, under contract with the National Aeronautics and Space Administration.

D.C. gratefully acknowledges financial support from the Fundaci\'on Andes. DZ acknowledges that this research was supported in part by the National Science Foundation under Grant No. PHY99-07949 during his visit to KITP, a Guggenheim fellowship, generous support from the NYU Physics department and Center for Cosmology and Particle Physics during his sabbatical there, NASA LTSA grant NNG05GE82G, NSF AST-0307482, and the David and Lucile Packard Foundation. JBH is funded by a Federation Fellowship from the Australian Research Council.

We particularly thank the referee, Michael Pohlen, for a very careful reading of the manuscript and extensive and helpful comments and suggestions.

%{\it Facilities:} \facility{Magellan:Clay ()}, \facility{Gemini:South ()}, \facility{VLT:Kueyen ()}, \facility{MMT ()}

\end{document}